\documentclass[11pt]{article}
\title{A Theory of Scattering Based on Free Fields}
\author{R.F. Streater\\Department of Mathematics\\King's College London}
\date{06 January 2010}
\begin{document}
\newtheorem{theorem}[equation]{Theorem}
\maketitle
\begin{abstract}
So far only quasifree fields have been shown to satisfy the Haag-Araki 
axioms for local algebras of observables; we show from a model in 1 + 1 dimensions that there can be representations in which two ingoing free particles 
produce a pair of out-going solitons with a positive probability, which can be computed. This happens when the experiment is designed to observe this 
outcome. It is proposed that the same idea will work in four dimensions.
\end{abstract}
\begin{center}
{\bf Contents}
\end{center} 
\begin{enumerate}
\item Introduction to Haag fields
\item Reduction to Abelian Multipliers
\item A Model in 1 + 1 dimensions
\item Some Remarks in Four Space-time Dimensions
\end{enumerate}

\section{Introduction to Haag Fields}

It has been extremely difficult to construct solutions to renormalisable quantum 
field theories that satisfy the Wightman axioms, in four space-time dimensions, 
except free fields and generalised free fields. It has been conjectured that 
quantum electrodynamics does not exist; only theories containing non-abelian
gauge fields, it is claimed, could exist and give a non-trivial S-matrix. Similar 
remarks apply to the $C^*$-algebraic systems of Haag and Araki.

The relation between the Wightman axioms and the $C^*$-algebras is not clear for a 
general Wightman theory, but for any free boson field a key result due to Slawny 
\cite{S} suggests a natural way to construct a set of local $C^*$-algebras which 
obey the Haag-Kastler axioms. Consider for example a free scalar quantised field of 
mass $m>0$. In any Lorentz frame, the free quantised field $\phi$ and its time 
derivative $\pi$ at constant time (say, time zero) can be smeared in the space 
variable with a continuous function of compact support, to get 
self-adjoint operators on Fock space. Thus
\begin{eqnarray}
\phi(g)&:=&\int \phi(0,{\bf x})g({\bf x})d^3x\\
\pi(f)&:=&\int\dot{\phi}(0,{\bf x})f({\bf x})d^3x
\end{eqnarray}
have well-defined exponentials, as do their sums; let $\cal H$ be the space of real 
solutions $\varphi(t,{\bf x})$ to the wave equation with initial values 
$\varphi(0,{\bf x})=f({\bf x})$ and $\dot{\varphi}(0,{\bf x})=
g({\bf x})$. This is a dense subspace of the one-particle space, a complex Hilbert 
space. The imaginary part of the scalar product, the symplectic structure of the 
classical field theory, is the Wronskian $B$ of the two solutions, the Lorentz 
invariant anti-symmetric bilinear expression
\begin{equation}
\hspace*{-.7in}B(\varphi_1,\varphi_2):=\int d^3x\left[f_1({\bf x})g_2({\bf x})-
g_1({\bf x})f_2({\bf x})\right]. 
\end{equation}
The expression 
\begin{equation}
B(\phi,\varphi) :=\phi(g)-\pi(f),
\end{equation}
the Wronskian between the quantised and the classical solution, is then 
self-adjoint. Segal uses the operators
\begin{equation}
W(\varphi):=\exp\{iB(\phi,\varphi)\},
\end{equation}
and these obey Segal's form of the Weyl relations for the commutation relations of a free 
quantised field:
\begin{equation}
W(\varphi_1)W(\varphi_2)=W(\varphi_1+\varphi_2)\exp\{-\frac{i}{2}
B(\varphi_1,\varphi_2)\}. \label{Segal-Weyl}
\end{equation}
Eq.(\ref{Segal-Weyl}) gives a product to the vector space defined by symbols 
$W(\varphi)$ as $\varphi$ runs over the symplectic space $\cal H$, irrespective of 
the representation by operators $W$. What Slawny \cite{S} did was to prove that the 
$^*$-algebra obtained by including this product has a unique $C^*$-norm; this is a 
norm on the algebra obeying $\|A^*A\|=\|A\|^2$. 

We define the Haag field as follows. Let $\cal O$ be a bounded closed set in 
${\bf R}^4$, of the form of the intersection of a forward and 
backward light cone. The cones themselves intersect in a two-dimensional ellipse. 
Let $f$ and $g$ be continuous functions of the points in the interior of the 
three-dimensional region spanned by this ellipse, and vanishing on the boundary. 
Then the local $C^*$-algebra 
${\cal A}(\cal O)$ is the completion, in the Slawny norm, of the Segal-Weyl
algebra generated by such $f$ and $g$. The algebra of all observables, $\cal A$, 
is then the completion of the inductive limit of all the local algebras. The 
algebra defined for an arbitrary connected open region of ${\bf R}^4$ is the 
completion of the union of all ${\cal A}(\cal O)$, $\cal O$ being a subset of the 
region.

This field nearly obeys the Haag-Kastler \cite{HK} axioms; Haag and Kastler assumed 
that the Poincar\'{e} group acted on ${\cal A}$ norm-continuously, which we do not. 
The free field satisfies one more, the split property of Doplicher and Roberts 
\cite{DR}. We use the notation ${\cal L}$ for the Poincar\'{e} group, which is the 
semi-direct product of the group of space-time translations, $x\mapsto x+a$, where 
$a$ is a real four-vector, and the Lorentz group $x\mapsto \Lambda x$. Thus 
$L=(a,\Lambda)$ will denote a general element of ${\cal L}$. Then the axioms we use 
are:
\begin{enumerate}
\item There is given an automorphism group $\tau_L$ of the Poincar\'{e} group; this 
maps ${\cal A}({\cal O})$ onto ${\cal A}(L{\cal O})$.
\item If two regions ${\cal A}_1$ and ${\cal A}_2$ are space-like separated, then 
the algebras ${\cal A}_1$ and ${\cal A}_2$ commute.
\item The vacuum representation: there exists a representation $R_0$ of ${\cal A}$, 
such that there is a unique vacuum state vector, the Poincar\'{e} group is 
continuously represented by unitary operators, and the spectrum of the energy is 
bounded below. 
\item The split property: if ${\cal O}_1^-\subset {\cal O}_2$ then there 
exists a sub-algebra ${\cal N}$ of type I such that ${\cal A}({\cal O}_1)
\subset{\cal N}\subset{\cal A}({\cal O}_2)$; by type I is meant that the weak 
closure in the vacuum representation is a von Neumann algebra of type I.
\end{enumerate}
Another possible axiom is Haag duality; this fails to hold in our model in 
one-plus-one dimensions and we shall not use it.

In their set-up, Haag and Kastler give the following explanation of superselection 
rules; charged states are not in the state-space containing the vacuum, but are 
states in some other representation $R$ of the algebra $\cal A$, which is not 
quasi-equivalent to $R_0$.  We mean the following by quasi-equivalence, which is 
equivalent to the usual definition for the algebras $\cal A$ arising in 
quantum field theory; let 
${\cal A}$ be such a $C^*$-algebra. A representation of ${\cal A}$, $\pi_1$ on a 
Hilbert space ${\cal H}_1$  is said to 
be {\em quasi-equivalent} to a representation $\pi_2$ on a Hilbert space 
${\cal H}_2$, if there exists an isometry $U: {\cal H}_2\rightarrow
{\cal H}_1$ such that 
\begin{equation}
U\pi_2(A)=\pi_1(A)U
\end{equation}
holds for all $A\in{\cal A}$. We say that an automorphism $\sigma$ of $\cal A$ is 
{\em spatial} in a representation $\pi$ of $\cal A$ on a Hilbert space $\cal H$ if 
there exists a unitary operator $U_\sigma$ on ${\cal H}$ such that
\begin{equation}
\sigma(A)=UAU^{-1}
\end{equation}
holds for all $A\in{\cal A}$. We say that $U$ implements the automorphism in this 
case. Most automorphisms are not spatial.

Haag and Kastler assume that the Poincar\'{e} automorphisms are spatial in $R$, 
and that the generator of time evolution also has positive spectrum. $R$ is related 
to the vacuum representation $R_0$ by an automorphism $\sigma$, $A\mapsto\sigma A$ 
of ${\cal A}$; this cannot be a spatial automorphism, since if it were, $R$ and 
$R_0$ would be equivalent. Clearly, the representation is given by 
\begin{equation}
R_\sigma(A)=R_0(\sigma A)\label{rep},
\end{equation}
as $A$ runs over ${\cal A}$; this acts on the Hilbert space containing the vacuum, 
but is not equivalent to the representation $R_0$, since the automorphism $\sigma$ 
is not implemented by a unitary operator. We do not expect $\sigma$ to commute with 
the space-time translations; 
thus, the automorphisms of $\cal A$, $\tau_a\sigma$, $a\in{\bf R}^4$, are not the same as $\sigma\tau_a$ 
in general. Haag showed that one might reveal the existence of Fermions, carrying a 
charge, by exploring the representations
\begin{equation}
R_n(A)=R_0(\tau_{_1}\sigma\circ\tau_{a_2}\sigma ...\tau_{a_n}\sigma A),
\end {equation}
which would define the $n$-particle states.
A little later, Doplicher and Roberts \cite{DR} generalised this idea; to get a 
representation of $\cal A$, one can make do with an {\em endomorphism} rather than 
an automorphism; one then gets a reducible representation of the algebra $\cal A$ 
by using eq\,(\ref{rep}). Doplicher and Roberts require that the endomorphism, call it $\sigma$, obeys $\sigma(I)=I$, so that the dual action of $\sigma$ on the states preserves normalisation. Thus Doplicher and Roberts use identity-preserving endomorphisms. The unitaries of the cummutant of the representation  
make up the gauge group. Starting with axioms similar to (1), ..., (4), they 
\cite{DR,H} find that the gauge group must be a compact Lie group. Now, this holds 
also for the free field algebra, though Doplicher and Roberts assumed that the 
given system was not the free field. They are stuck, in that no interacting 
Wightman theory in four dimensions has yet been constructed.

In this paper, we start with the free field as in \cite{S07}, and try to find what 
endomorphisms give rise to new states. We note that it is not obvious that the 
Lorentz group should be implemented in $R$ even if the space-time translations
are; more, the space-time group might acquire non-abelian multipliers. In Sect (2) 
we show that if every one-parameter space-time translation group with a time-like 
direction has 
spectrum that is bounded below, then the four-dimensional 
translation group is represented by unitaries which have multipliers in the centre 
of $R({\cal A})^{\prime \prime}$. This proof uses Borchers's theorem \cite{B} in 
the form proved in Bratteli and Robinson \cite{BR}; it arose from a discussion 
with G. Morchio. We are then reduced to the suggestion of several authors, that 
the space-time group might be represented with multipliers in the centre.

In Sect. (3) we study the case of a free massless field in 1 + 1 dimensions, 
following \cite{SW}. This model has been further developed by Ciolli \cite{C}. 
We show that a soliton pair of states with opposite charges does lie in Fock space, 
and converges $*$-weakly to an out-going pair in a new representation. The pair is 
created from a state in Fock space by the 
very act of asking the question, is a pair present at $t=\infty$? 
 
In Sect.(4) we
suggest a programme that might lead to similar results in four space-time dimensions.
                                 
\section{Reduction to Abelian Multipliers}

It is usually required that the endomorphism, denoted by $\sigma$ above, should be 
such that the Poincar\'{e} group be spatial in the representation $R_\sigma$. 
However, with particles of zero mass, it might not be true. In any case, we shall 
just assume that space-time translations are symmetries in $R_\sigma$; that is, are 
each given by an isometric operator with transition probabilities that are 
measurable functions of the group  parameters; then Wigner's analysis can be applied. 
Now, $R_\sigma$ is reducible if $\sigma$ is not an automorphism; thus the commutant 
$R_\sigma({\cal A})^\prime$ of the representation contains non-commuting unitaries, 
and so possible multipliers of the group ${\bf R}^4$ might be non-abelian 
\cite{RFS1,RFS2}. It is well known that a one-parameter group of automorphisms, 
if spatial in a representation, has only trivial multipliers \cite{J}. It has been 
suggested that conditions might be such that the multiplier is abelian. Indeed, 
there does exist a natural condition which ensures this. 
\begin{theorem}
Let $\cal A$ be a $C^*$-algebra and $\tau_a$ be a group action of ${\bf R}^4$ by 
automorphisms. Let $A\mapsto R(A)$ be a representation of $\cal A$ such that 
the group action is weakly measurable.
Suppose that for each time-like one-parameter subgroup of ${\bf R}^4$, the 
automorphisms are implemented (in the representation $R$) by a continuous 
one-parameter unitary group, whose self-adjoint generator is bounded below. 
Then the group ${\bf R}^4$ is projectively represented by unitary operators with 
abelian multipliers.
\end{theorem} 
Proof. Borchers's theorem \cite{B} was modified by Bratteli and Robinson \cite{BR} 
to the form: 
let $\cal A$ be a $C^*$-algebra on a separable Hilbert space, $\tau_t$  a 
one-parameter group of automorphisms of 
$\cal A$, implemented by the continuous one-parameter unitary group $t\mapsto U(t)$. 
Then there exists a continuous unitary group $t\mapsto V(t)$ in the weak closure of 
$\cal A$ which implements $\tau_t$. 

We apply this result to four independent one-parameter timelike one-parameter 
groups of space-time translations. Choose four linearly independent time-like 
vectors $a_i$, $i=1,\ldots 4$. The generators are bounded below, and so can be 
replaced by unitary operators in the weak closure. The multipliers, which are 
expressed as 
\begin{equation}
\omega(a_i,a_j)=U(a_i)U(a_j)U(a_i+a_j)^{-1},\label{mult}
\end{equation}
shows that for each pair of our four time-like vectors we have $\omega(a_i,a_j)$ 
lying in $R({\cal A})^{\prime\prime}$; 
but these multipliers also lie in $R({\cal A})^\prime$, so must lie in the centre. 
For any $\lambda\in{\bf R}$ we may implement $x\mapsto x+\lambda a_i$ by 
$U(\lambda a_i):=U(a_i)^\lambda$, for any measurable choice of the branch. Since 
$R(\cal A)^{\prime\prime}$ is a von Neumann algebra, we have that 
$U(\lambda a_i)\in R(\cal A)^{\prime\prime}$. Now, these group 
elements generate the group ${\bf R}^4$, and for any translation $y$ we 
have unique $\lambda_i$, $i=1,\ldots,4$ such that $y=\lambda_j a_j$; we may define 
$U(y):=U(\lambda_1 a_1)\ldots U(\lambda_4 a_4)$ which implements the automorphism 
$y$ and lies in $R(\cal A)^{\prime\prime}$. We prove the theorem by using 
eq.(\ref{mult}) for any two elements of ${\bf R}^4$, which shows that 
$\omega(y_1,y_2)$ lies in the centre.

\section{A Model in One-Plus-One Dimensions}

The existence of Wightman theories with interaction in  $1+1$-dimensions \cite{GJ} means that it has 
not been necessary to consider our idea in this case; however, in view of the difficulty, if not 
the impossibility, of there existing a Wightman theory in four space-time dimensions, it is 
worth while pointing out the following model.

Consider the Wightman theory of a scalar massless free field $\phi(x,t)$ in $1+1$ dimensions. This does 
not exist as a Wightman theory, but the system given by its space-time derivatives, $\phi_\mu:=\partial_\mu\phi$, does. We take this derivative, 
$\mu=0,1$, to define the observable Wightman fields. The smeared fields $\phi_\mu$ at time zero, obey 
a form of the CCR which can be written in Segal form. We \cite{SW} get a Haag field, and show that 
it obeys axioms 1, 2 and 3. We consider new representations of the form
\begin{eqnarray}
\partial_x\phi_\sigma&=&\partial_x\phi+\partial_x\varphi\\
\partial_t\phi_\sigma&=&\partial_t\phi+\varpi.
\end{eqnarray}
Here, $\varphi$ and $\varpi$ are real-valued smooth functions, and such that $\partial_x\varphi$ and 
$\varpi$ have compact support. It is known that the representation obtained this way is equivalent to 
the Fock representation if and only if the classical solution determined by the initial values 
$\varphi,\varpi$ lies in the one-particle space.
We showed \cite{SW} that there exists a two-parameter family of superselection rules, labelled by 
``charges'' $Q,Q^\prime$ say; these can be any pair of real numbers. If they are both zero, then the 
automorphism is spatial in the free Fock representation. The set of $\varphi$ allowed consists of 
functions such that $\partial_x\varphi\in{\cal D}$, and the set of $\varpi$ is $\cal D$ itself, 
Schwartz space;  this can lead to states not in Fock space. Two representations with different 
values of either $Q$ or $Q^*$ are inequivalent; it is thus reasonable to put the discrete topology 
on the set ${\bf R}^2$. The dual of this topological space is thus the compact gauge group
$U(1)\times U(1)$.

Consider, for example, the choice of $Q=1,\;Q^\prime=1$. A general solution to the wave equation can be 
written as the sum of a left-going and a right-going wave:
\begin{equation}
f(x,t)=f_{_L}(x+t)+f_{_R}(x-t).
\end{equation}
We see that a left-going wave can have $Q=1$ and $Q^\prime=1$ if $f_{_R}=0$ and $f_{_L}=\varphi$, 
$\varpi=\partial_x\varphi$,
where $\varphi(x)=1$ if $x$ is sufficiently large, and $\varphi(x)=0$ for $x$ sufficiently negative. 
It follows that there is a state in Fock space, with $\varphi$ consisting of a right-moving 
positive bump to the right of space, with 
$Q=-1$ and $Q^\prime=-1$, and a left-moving negative bump to the left of space, with $Q=1$ and 
$Q^\prime=1$. Let $F(x,t)$ be classical solution with these properties. Then the automorphism is 
implemented by the unitary operator $W(F)=\exp\{i\left(\phi(\dot{F})-\pi(F)\right)\}$.

As time goes by, these solitons move as out-going free particles. There is a non-zero probability $P$ that a 
given two-particle 
state $\left.|2\right>$ in Fock space will lead to this configuration:
\begin{equation}
P=|\left<2|W\Psi_0\right>|^2 >0.
\end{equation}

It is clear that if we look for the free particles, we will see them; no new 
particles are produced. The charged particles are produced by the setting-up of 
the procedure to see them.

Further work on this model was done by Ciolli \cite{C}. He proved using 
Roberts's net cohomology \cite{R} that all possible superselection rules 
were found in \cite{SW}.
 
\section{An Attempt in Three + One Dimensions}

The electromagnetic field obeys the Maxwell equations
\begin{eqnarray}
{\rm div}\,{\bf E}&=&\rho\\
{\rm div}\,{\bf B}&=&0\\
\partial_t{\bf B}&=&-{\rm curl}\,{\bf E}\\
\partial_t{\bf E}&=&{\rm curl}\,{\bf B}+{\bf j}
\end{eqnarray}
The free-field arises when $\rho$ and ${\bf j}$ vanish; the classical electromagnetic wave is described 
by a transverse free ${\bf E,B}$. That is, ${\bf E}$ and ${\bf B}$ are both orthogonal to the momentum 
of the wave. There are two states, labelled by the polarisation, for each momentum. The set of such 
solutions form a real Hilbert space, with a symplectic form and a complex structure. The action of the 
Poincar\'{e} group is unitary, the representation being of mass zero and helicity $\pm 1$. The three 
components of ${\rm curl}\,{\bf E}$ are transverse, even when $\rho$ is not zero. For, the distribution 
${\rm curl}\,{\bf E}$ has three components. The $x$-component is $\partial_zE_y-\partial_yE_z$; thus, 
${\rm curl}\,{\bf E}$, smeared with the three-vector ${\bf f}$, is the space of operators
\[
{\rm curl}\,{\bf E}.(\bf f)={\bf E}.{\rm curl}\,{\bf f}\]
whereas the longitudinal part of the field is of the form ${\bf E}.\nabla g$.
Since the set ${\rm curl}.{\bf f}$ is disjoint from the set of $\nabla g$ except for $0$, we have shown 
that ${\rm curl}\,{\bf E}$ is transverse.

Smeared with test functions ${\bf f}$ in ${\cal D}({\bf R}^3)$, the
functions ${\rm curl}\,{\bf f}$ are dense in the one-particle space. We 
define the local $C^*$-algebra ${\cal A}(\cal O)$ using Slawny's theorem, using test-functions in 
${\cal D}({\cal O})$. The global $C^*$-algebra $\cal A$ is the completion of the union of all such 
algebras for bounded regions in space-time. Let $R$ be the relativistic Fock representation of the 
transverse electromagnetic field. 

We seek an identity-preserving endomorphism $\sigma$ of $\cal A$ so that the representation obtained by $R_\sigma(A)=
R(\sigma(A))$ is disjoint from the representation $R_{_0}$. More, we need that the space-time 
automorphisms of ${\cal A}$ should be spatial in $R_\sigma$, and that any  one-dimensional time-like 
translation group should be continuous, and that its generator should be bounded below. The dynamics 
of the operators in $R_\sigma$ is given by the free automorphism group of the free field. However it 
is not a trivial dynamics, so we hope. The Hamiltonian is not 
a bounded operator, and neither are the field operators. So these are not in the $C^*$-algebra, and 
their algebraic properties might not be preserved if we change to an inequivalent representation. 
The commutator of these gives the time 
evolution of the field operator. However, the Lie algebra of such commutators might not be 
preserved under the endomorphism: 
there might be new terms, an induced interaction. This is due to the anomalies that arise in commutators.
Another possibility, which changes the equations of motion, is to change coordinates of space-time by a 
smooth but non-linear map. This might lead to a new representation, but it is not clear that the space-time 
translations would be spatial in the new representation. 

Leyland and Roberts \cite{LR} have used the theory of sheaf cohomology to study the possible two-cocycles 
of some free classsical fields in Minkowski space. They conclude that for the scalar Klein-Gordon real field, 
the two-cohomolgy group is trivial, while for the free maxwell field there is a two-parameter family 
of two-cocycles, 
labelled by electric and the magnetic charge. They also showed that the classical four-potential, $A_\mu$, 
obeying the subsidiary condition $\partial^\mu A_\mu=0$ and the wave equation $(\partial_0^2-\Delta)A_\mu=0$, showed a 
one-parameter family of electric charges. It is not clear from their remarks that this holds in the 
quantum case, which requires non-commuting operators for the fields; however, it does hold. As we did in 
1 + 1 dimensions, we can add this classical solution to the free quantised field, to generate an 
automorphism of the free field algebra. When we add a cocycle which is not a coboundary, we get 
a new representation. Leyland and Roberts do not consider the condition that the Maxwell field should 
be transverse, nor the requirement that the new representations
found should have energy bounded below. The latter condition can be satisfied if we require that the 
solution should extend to the point at infinity, as in the methods described by Ward and wells \cite{WW}.
This is possible only for a subset of the solutions, namely, those with integer charge. Thus, the problem 
with continuous charge can be solved in this way. We can remove the occurrence of magnetic charge 
by requiring the existence of a potential $A_\mu$. However, this work leads to sectors with zero 
mass, since there is no mass-parameter in the model. This leads to doubts that it is an electron.

Of interest is the model of Prasad and Sommerfield \cite{PS}. They explicitly construct a smooth 
solution of a free massive boson field in a non-abelian gauge field, and the electromagnetic part 
of the gauge field has a magnetic pole as well as an electric pole. The energy of the solution is 
finite. The rigorous treatment \cite{WW} concerns the analytic continuation from Minkowski to 
Euclidean space ${\bf R}^4$. It mostly assumes that the Euclidean gauge field is dual or anti-dual 
${\bf E}=\pm i{\bf H}$, though the book also deals with some non-self-dual electromagnetic fields. 
Donaldson \cite{D} has pointed out that 
in four dimensions, in the Euclidean formulation, and in the case of self-dual electromagnetic tensors, 
the second sheaf cohomolgy group is non-trivial. He remarks that this would furnish ${\bf R}^4$ with 
new differential structures.
From the point of view of the second quantised theory, the $C^*$-algebra of the electromagnetic field 
reveals the charge in its equations of motion in the corresponding representation.

The book \cite{WW} deals with the classical version of this problem. However, 
for linear fields, this is close to the quantum version, as we saw in \cite{SW}; 
we use the 
classical solution to get the displaced Fock representations.
Further, the non-linearity of the gauge field in classical field theory can sometimes be linearised by 
a suitable change of coordinates. 
The representations obtained by smooth invertible change of coordinates are generally spatial in Fock space;
they would produce unstable particles instead of superselected states. 
The Euclidean approach of Symanzik \cite{Sy} and Nelson 
\cite{N} might be the way to proceed; a coordinate change in Euclidean variables could lead to 
the correct version of the relation between the Fock and non-Fock representations.  

In 4 + 1 dimensions, Vasilliev has shown that a four-dimensional change of coordinates leads us to 
the soliton, which obeys a Dirac equation.

\end{document}